\documentclass[aps,preprint,nofootinbib,tightenlines,superscriptaddress]{revtex4}
\usepackage{epsfig}
\usepackage{amsmath}
\usepackage{graphicx}
\usepackage{multirow}   

\newcommand{\beq}{\begin{equation}}
\newcommand{\eeq}{\end{equation}}

\newcommand{\bold}{\textbf}

\begin{document}

\title{ Fully-heavy tetraquark spectra and production at hadron colliders}

\author{Ruilin Zhu}

\email{rlzhu@njnu.edu.cn}
\affiliation{Department of Physics and Institute of Theoretical Physics,
Nanjing Normal University, Nanjing, Jiangsu 210023, China}
\affiliation{Nuclear Science Division, Lawrence Berkeley National Laboratory, Berkeley, CA 94720, USA}

\date{\today}

\begin{abstract}
Motivated by the observation of exotic structure around 6900MeV in the $J/\psi$-pair mass spectrum  using proton-proton collision data
by the LHCb collaboration, we study the spectra of fully-heavy tetraquarks within Bethe-Salpeter equation and Regge trajectory relation. The $X(6900)$ may be
explained as a radially excited state with quark content $cc\bar{c}\bar{c}$ and spin-parity $0^{++}(3S)$ or $2^{++}(3S)$  or an orbitally
excited $2P$ state. New $cc\bar{c}\bar{c}$
structures around
6.0GeV, 6.5GeV, and 7.1GeV are predicted together. Other
$bb\bar{b}\bar{b}$  and $bc\bar{b}\bar{c}$ structures which may be experimentally prominent are discussed.  On the other hand, the fully-heavy
S-wave tetraquark production at hadron colliders are investigated and
their cross sections are obtained.
\end{abstract}

\maketitle

\section{Introduction}
\label{SEC:Introduction}

Color confining is a long-standing open question since the invention of Quantum Chromodynamics (QCD).
To understand its nature, one phenomenological way is to establish the hadron spectroscopy.
In naive quark model, meson is composed of a quark-anqiquark pair while baryon is composed of three quarks.
However, one could also imagine in general that there would be the possibility of Glueball with parton content
$(gg, ggg,...)$ and multi-quark states with parton content $(qq\bar{q}\bar{q}, qqqq\bar{q},...)$ since
they do not violate the principle of QCD color confining. There is great progress in the search of multi-quark states.
Many new exotic structures beyond the
naive quark model have been observed since the discovery of $X(3872)$ in 2003~\cite{Choi:2003ue,Acosta:2003zx}, for a  recent
review
of these new structures, see
Refs.~\cite{Guo:2017jvc,Olsen:2017bmm,Liu:2019zoy,Brambilla:2019esw,Yang:2020atz}).

Very recently, the LHCb collaboration have reported the discovery of one exotic structure around 6900MeV in the
invariant mass spectrum of $J/\psi$ pairs at the Large Hadron Collider~\cite{Aaij:2020fnh}.
This is a possible candidate for a fully charm tetraquark and has the quark content $cc\bar{c}\bar{c}$.
Its mass and decay width are determined from two model scenarios
\begin{align}
 m_{X(6900)}&=6905\pm 11\pm7{\rm MeV},\nonumber\\
 \Gamma_{X(6900)}&=80\pm19\pm33{\rm MeV},\nonumber
\end{align}
for model-I with no interference with the non-resonance single parton
scattering continuum;
\begin{align}
 m_{X(6900)}&=6886\pm 11\pm11{\rm MeV},\nonumber\\
 \Gamma_{X(6900)}&=168\pm33\pm69{\rm MeV},\nonumber
\end{align}
for model-II with interference with the non-resonance single parton
scattering continuum.

Theoretically, many phenomenological models have investigated  the fully heavy tetraquark states in Refs.~\cite{Iwasaki:1975pv,Chao:1980dv,
Ader:1981db,Badalian:1985es,Heller:1985cb,Lloyd:2003yc,Barnea:2006sd,Vijande:2009kj,Berezhnoy:2011xn,
Heupel:2012ua,Wu:2016vtq,Chen:2016jxd,Karliner:2016zzc,Bai:2016int,Wang:2017jtz,
Richard:2017vry,Anwar:2017toa,Debastiani:2017msn,Richard:2018yrm,Esposito:2018cwh,Wang:2018poa,
Liu:2019zuc,Wang:2019rdo,Bedolla:2019zwg,Chen:2020lgj,Deng:2020iqw,Lundhammar:2020xvw,
liu:2020eha,Yang:2020rih,Wang:2020gmd,Garcilazo:2020acl,Albuquerque:2020hio,Giron:2020wpx,Sonnenschein:2020nwn,Maiani:2020pur,
Richard:2020hdw,Wang:2020wrp,Chao:2020dml,Maciula:2020wri,Karliner:2020dta,Jin:2020jfc,Chen:2020xwe,Wang:2020dlo,Dong:2020nwy,Ma:2020kwb,
Feng:2020riv,Zhao:2020nwy,Gordillo:2020sgc,Weng:2020jao,Jamil:2020ull,Zhang:2020xtb}. In these studies, most of them verified the existence of stable
states with four heavy quarks and predicted the ground states of fully heavy tetraquarks below $X(6900)$. To explain $X(6900)$, some of them
assigned it as a radially or orbitally excited state, such as Refs.~\cite{Chen:2020xwe,Zhao:2020nwy}.
Note that there is also a CMS report on possible structure in $(18,19)$GeV region~\cite{CMS-Yi}, hinting a potential fully bottom tetraquark.
 The discovery of a fully heavy
tetraquark shall definitely deepen our knowledge of hadron structure.

For a $cc\bar{c}\bar{c}$ system, the spin-parity $J^{PC}$ can be  $0^{++}, 1^{+-}$ and $2^{++}$ for  S-wave states while
$0^{-+}, 1^{-+}, 1^{--}, 2^{-+}, 2^{--}$ and $3^{--}$ for  P-wave states. Considering the $X(6900)$ directly decays into a pair
of $J/\psi$, the allowed quantum numbers of spin-parity are $0^{++}$ and $2^{++}$ for  S-wave states while
$0^{-+}$, $ 1^{-+}$ and $2^{-+}$ for  P-wave states.
In this paper, we will argue the $X(6900)$ is more likely to be an excited state rather than a ground state of fully charm tetraquark
within the Bethe-Salpeter equation and Regge trajectory relation.
Then we employ the formulae to study the spectra of other fully heavy tetraquarks such as $T_{bb\bar{b}\bar{b}}$
 and $T_{bc\bar{b}\bar{c}}$.
  On the other hand, we will study the production mechanism of fully heavy S-wave tetraquarks at hadron colliders.
  The color-singlet and color-octet configurations will be discussed.
 We will employ the nonrelativistic QCD (NRQCD)~\cite{Bodwin:1994jh} to predict the cross section of fully heavy tetraquarks
 $T_{cc\bar{c}\bar{c}}$, $T_{bb\bar{b}\bar{b}}$
 and $T_{bc\bar{b}\bar{c}}$.

The rest of this paper is organized  as follows.  The  spectra of fully heavy tetraquarks $T_{cc\bar{c}\bar{c}}$, $T_{bb\bar{b}\bar{b}}$
 and $T_{bc\bar{b}\bar{c}}$ are given by Bethe-Salpeter equation in Sec.~II.  Their spectra are also discussed based
 on the Regge trajectories in Sec.~III. The total cross sections  are
presented in Sec.~IV. We also phenomenologically  discuss how to hunt for the possible $X(6900)$ partners.
 The differential cross section is investigated at low transverse momentum in Sec.~V.
In the end we give a brief summary.

\section{The Bethe-Salpeter equation for a diquark-antidiquark state}
\label{SEC:spectra0}
It is complicated to study a four-body interaction. In this section, we will study the spectra of diquark-antidiquark states in quasi two-body interactions. In general the Bethe-Salpeter wave functions for a $0^{++}$ and $2^{++}$ diquark-antidiquark states are
\begin{equation}
\chi'\left(x_{1}, x_{2}, p\right)=\left\langle 0\left|T [{\cal D}_0\left(x_{1}\right) {\cal D}_0\left(x_{2}\right)]\right| T_{4Q}(p)\right\rangle,
\end{equation}
\begin{equation}
\chi\left(x_{1}, x_{2}, p\right)={\cal P}_{\mu\nu}\left\langle 0\left|T [{\cal D}_1^{\mu}\left(x_{1}\right) {\cal D}_1^{\nu}\left(x_{2}\right)]\right| T_{4Q}(p)\right\rangle,
\end{equation}
\begin{equation}
\chi^{\mu\nu}\left(x_{1}, x_{2}, p\right)=\left\langle 0\left|T[ {\cal D}_1^{\mu}\left(x_{1}\right) {\cal D}_1^{\nu}\left(x_{2}\right)]\right| T_{4Q}(p)\right\rangle.
\end{equation}

The diquark momenta can be written as
\begin{eqnarray}
p_1 &=&  \alpha_1 \,p+q,\\\
p_2 &=&  \alpha_2\, p-q,
\end{eqnarray}
where $p=p_1+p_2$ is the tetraquark momentum; $q$ is a half of the relative momentum between the
diquark pair; $\alpha_1=m_1/(m_1+m_2)$ and $\alpha_2=m_2/(m_1+m_2)$ with the different diquark masses $m_1$ and $m_2$.

In the following, we will take the $0^{++}$ scalar diquark-antidiquark as an example  and discuss its  Bethe-Salpeter equation. The Bethe-Salpeter wave function for the $0^{++}$ scalar diquark-antidiquark can be rewritten as
\begin{equation}
\chi'\left(x_{1}, x_{2}, p\right)=e^{-i p \cdot X} \int \frac{d^{4} q}{(2 \pi)^{4}} e^{-i q \cdot x} \chi_{p}(q),
\end{equation}
where two coordinators are $X=\alpha_1 x_1+\alpha_2 x_2$ and $x=x_1-x_2$.

The Bethe-Salpeter equation for  $0^{++}$ scalar diquark-antidiquark state is
\begin{equation}
\chi_{p}(q)=S_{{\cal D}_0}\left(p_{1}\right) \int \frac{d^{4} k}{(2 \pi)^{4}} K(p, q, k) \chi_{p}(k) S_{{\cal D}_0}\left(p_{2}\right),\label{BSE}
\end{equation}
where the diquark propagator is
\begin{equation}
S_{{\cal D}_0}\left(p_{i}\right)=\frac{i}{p_i^2-m_i^2+i\epsilon}= \frac{i}{\left(\alpha_i p-(-1)^i q_\parallel\right)^2-\omega_i^2+i\epsilon},
\end{equation}
with $\omega_i=\sqrt{m_i^2+|\bold q|^2}$.
The above Bethe-Salpeter equation is defined in four dimensional space-time.
However, it can be reduced into a three dimensional equation when consider
the so-called ``instantaneous approximation''\cite{Chang:2004im,Chang:2006tc,Ding:2021dwh}. The Bethe-Salpeter kernel does not
depend on the time component of the diquark relative momentum,
\begin{equation}
K(p, q, k)|_{\bold p=0}\simeq M^2 V(\bold q, \bold k)= M^2  V(\bold q-\bold k),
\end{equation}
where the on shell condition is $p^2=M^2$ and the kernel $V$  is the potential.

We can rewrite the relative momentum $q$ into two components,
\begin{equation}
q^\mu=q^\mu_\parallel+q^\mu_\perp,
\end{equation}
where $q^\mu_\parallel=\frac{(p\cdot q)p^\mu}{M^2}$ and $q^\mu_\perp=q^\mu-q^\mu_\parallel$ are a parallel component and
an orthogonal component to the tetraquark state momentum $p$ with $p^2=M^2$ respectively.

Four simplicity, we consider the identical components and then $\alpha_1=\alpha_2=1/2$ and $\omega_1=\omega_2=\omega$. After integrate $q_\parallel$ out on both sides of Eq.~(\ref{BSE}), we can  obtain
\begin{equation}
(M-2\omega) \chi_{p}(q_\perp) =\tau(M,\omega)\eta(q_\perp),
\end{equation}
where $\tau(M,\omega)=\frac{M^2}{\omega (M+2\omega)}$ and the instantaneous Bethe-Salpeter wave function is
\begin{align}
\chi_{p}(q_\perp)=\int\frac{d q_\parallel}{2\pi}\chi_{p}(q)=\int\frac{d q_\parallel}{2\pi}\chi_{p}(q_\parallel,q_\perp),\\
\end{align}
and the integration
\begin{align}
\eta_{p}(q_\perp)=\int\frac{d^3 k_\perp}{(2\pi)^3}V(q_\perp-k_\perp)\chi_{p}(k_\perp).\\
\end{align}

The diquark components are heavy and then we can use the nonrelativistic approximation with $m_i\to \infty$.
Then the energy factor $\tau(M,\omega)=\frac{M^2}{\omega (M+2\omega)}\to 1$ and the Bethe-Salpeter
equation turns to the Schr\"{o}dinger equation.
\begin{equation}
\left(-\frac{\hbar^{2}}{2 \mu_{R}} \nabla^{2}+V_{{\cal{D} }{\cal{\bar{D}}}}(r)\right) \chi_p(\mathbf{r})=E \chi_p(\mathbf{r}),
\end{equation}
where $\mu_R=m_{\cal D}/2$ is the reduced mass.
The potential is assumed as  a long-ranged linear confining potential, a short-ranged one gluon exchange potential,
and a spin-orbital splitting term~\cite{Eichten:1994}. For S-wave states with orbital angular momentum $\ell=0$, the potential in coordinate space is chosen  as~\cite{Cao:2012du}
\begin{align}
V_{{\cal{D} }{\cal{\bar{D}}}}(r)=-\frac{4}{3}\frac{\alpha_s}{r}+\lambda r+ V_S (r),
\end{align}
where the spin dependent term is
\begin{align}
 V_S (r)=\frac{32\pi \alpha_s}{9m_{\cal D}^2}\left(\frac{\delta}{\sqrt{\pi}}\right)^3 e^{-\delta^2 r^2} S_{\cal D}\cdot S_{\cal \bar D},
\end{align}
and  we have spin matrix element
\begin{align}
 \langle S_{{\cal D}_i}\cdot S_{{\cal \bar D}_j}\rangle=\frac{1}{2}\left(S(S+1)-S_i(S_i+1)-S_j(S_j+1)\right).
\end{align}

The following parameter values are adopted as: the scalar and axial-vector diquark masses $m_{{cc}_0}=2.95GeV$ and
$m_{{cc}_1}=3GeV$, $m_{{bc}_0}=6.2GeV$, $m_{{bc}_1}=6.25GeV$, $m_{{bb}_0}=9.4GeV$, $m_{{bb}_1}=9.45GeV$,  $\lambda=0.1468(GeV)^2$, $\delta=1.1384GeV$. The Schr\"{o}dinger equation can be numerically evaluated by the diagonalization method of the matrix and the stationary eigenvalues can be obtained when we set the
 number of steps, $N\sim 1000$. We then list our results in the Tab.~\ref{MassT4Q}.
\begin{table}\caption{Predictions of the masses ({\rm MeV}) of S-wave fully heavy $T_{4Q}(n S)$ tetraquarks. Only $0^{++}$
and $2^{++}$ are considered for $T_{bc\bar{b}\bar{c}}$. The uncertainty is from the coupling constant $\alpha_s=0.35\pm0.05$.}
{\small \begin{tabular}{|c|c|c|c|c|c|}\hline\hline
$T_{4Q}(nS)$ states  & $J^P$ & Mass($n=1$)& Mass($n=2$)& Mass($n=3$)& Mass($n=4$)\\\hline
$T_{cc\bar{c}\bar{c}}$ &$0^{++}$ &$6055_{-74}^{+69}$ &$6555_{-37}^{+36}$&$6883_{-27}^{+27}$&$7154_{-22}^{+22}$
\\
&$2^{++}$ &$6090_{-66}^{+62}$ &$6566_{-35}^{+34}$&$6890_{-26}^{+27}$&$7160_{-22}^{+21}$
\\\hline
$T'_{cc\bar{c}\bar{c}}$ &$0^{++}$ &$5984_{-67}^{+64}$ &$6468_{-35}^{+35}$&$6795_{-26}^{+26}$&$7066_{-22}^{+21}$
\\\hline
$T_{bc\bar{b}\bar{c}}$ &$0^{++}$ &$12387_{-120}^{+109}$ &$12911_{-51}^{+48}$&$13200_{-36}^{+35}$&$13429_{-30}^{+29}$
\\
&$2^{++}$ &$12401_{-106}^{+117}$ &$12914_{-49}^{+49}$&$13202_{-36}^{+35}$&$13430_{-29}^{+29}$
\\\hline
$T'_{bc\bar{b}\bar{c}}$ &$0^{++}$ &$12300_{-117}^{+106}$ &$12816_{-50}^{+48}$&$13104_{-35}^{+35}$&$13333_{-29}^{+29}$
\\\hline
$T_{bb\bar{b}\bar{b}}$ &$0^{++}$ &$18475_{-169}^{+151}$ &$19073_{-63}^{+59}$&$19353_{-42}^{+42}$&$19566_{-35}^{+33}$
\\
&$2^{++}$ &$18483_{-168}^{+149}$ &$19075_{-62}^{+59}$&$19355_{-43}^{+41}$&$19567_{-35}^{+33}$
\\\hline
$T'_{bb\bar{b}\bar{b}}$ &$0^{++}$ &$18383_{-167}^{+149}$ &$18976_{-62}^{+59}$&$19256_{-42}^{+43}$&$19468_{-34}^{+34}$
\\\hline
\hline\hline
\end{tabular}}\label{MassT4Q}
\end{table}

In principle, the wave functions of the tetraquark include the space, flavor, spin and color parts. Considering
the fact that the $QQ$ diquark has symmetrical flavor and space wave functions, the axialvector diquark is
in the color-triplet configuration, while the scalar diquark is in the color-sextet configuration. Thus
we can imply that the  $T_{4Q}(0^{++},2^{++})$ states are more stable than the $T'_{4Q}(0^{++})$ state.

\section{Regge trajectories for the fully heavy tetraquark spectra}
\label{SEC:spectra}
The $X(6900)$ is around 700MeV above the $J/\psi J/\psi$ threshold, thus it is not likely
to be a ground state of fully charm tetraquark. Actually many literatures have predicted the
ground state of fully charm tetraquark around 6GeV, see literatures such as
\cite{Heupel:2012ua,Karliner:2016zzc,liu:2020eha,Chen:2020xwe,Zhao:2020nwy} or
Tab. VIII in Ref.~\cite{Gordillo:2020sgc}.
In the following, we will use the Regge trajectories to study the excited fully heavy tetraquarks
and attempt to understand the $X(6900)$ spectra.

In Regge  theory, all hadrons  (stable or unstable baryons and mesons) can be associated
with Regge poles that move in the  angular momentum plane as a function of hadron mass~\cite{Chew:1962eu}.
Later developments indicate this relation in $(J, M^2)$ plane is approximately linear
\begin{align}
J=\alpha M^2+\alpha_0,
\end{align}
where $J$ is the spin quantum number and $M$ is the hadron mass.
 $\alpha$ is the slope and $\alpha_0$ is the intercept, both of which are
 model parameters and different for different baryons and mesons.
On the other hand, it is convenient to
construct the hadron Regge trajectories in $(n_r, M^2)$ plane
~\cite{Ebert:2009ub,Ebert:2009ua,Ebert:2011jc,He:2020jna}
\begin{align}
n_r=\beta M^2+\beta_0,
\end{align}
where $n_r=n-1$ with the radial quantum number $n$.
The slope $\beta$ and the intercept $\beta_0$ are also the free parameters and dependent on certain hadron.

One interesting remark is that the slopes decrease when much heavier quark gets in the hadron.
For the hadrons with identical constituent quark content, the slopes are at the same order.
From the global fits of spectra of all known meson data  and higher excited states from QCD-motivated relativistic
quark potential model in Refs.~\cite{Ebert:2009ub,Ebert:2009ua,Ebert:2011jc}, $\alpha(q\bar{q})\subset[0.828,1.336]\sim\alpha(q\bar{s})\subset[0.780,0.964]
>\alpha(s\bar{s})\subset[0.684,0.729]>\alpha(q\bar{c})\subset[0.489,0.557]\sim\alpha(s\bar{c})\subset[0.463,0.497]>\alpha(q\bar{b})
\subset[0.243,0.288]\sim\alpha(s\bar{b})\subset[0.241,0.290]$, and $\beta(q\bar{q})\subset[0.679,0.916]
>\beta(s\bar{s})\subset[0.559,0.597]>\beta(q\bar{c})\subset[0.339,0.378]>\beta(s\bar{c})\subset[0.309,0.336]
>\beta(q\bar{b})\subset[0.172,0.183]
\sim\beta(s\bar{b})\subset[0.169,0.177]$. For charmonium, $B_c$ and bottomonium systems, the  fitted slopes are
$\alpha(c\bar{c})\subset[0.414,0.493]>\alpha(c\bar{b})\subset[0.242,0.298]\sim\alpha(b\bar{b})\subset[0.184,0.267]$ and
$\beta(c\bar{c})\subset[0.287,0.325]>\beta(c\bar{b})\subset[0.172,0.190]\sim\beta(b\bar{b})\subset[0.151,0.178]$.
Based on these fits, one could expect the slopes  are approximately equal for hadrons with identical heavy quark content but with different spin-parity.

The other remark is that the parameters $\alpha_0$ and $\beta_0$ are not unpredictable. At least, its value can be well estimated
by the ground states of hadron due to
 \begin{align}
\alpha_0=-\alpha M^2(J=J_{min})+J_{min},  ~~~ \beta_0=-\beta M^2(n=1).\label{intercep}
\end{align}

In the following, we will update the fitting of the  slope and intercept for heavy quarkonium and $B_c$ meson systems. Compared to
the fitting in Refs.~\cite{Ebert:2011jc}, we increase the weight of experimental data of heavy quarkonium and $B_c$ meson spectra and
decrease the weight of unobserved states such as discarding unobserved higher excited states ($J>3;n_r>3$) predicted from potential models.
 We adopt Chi-square fit and the Chi-square goodness of fit
 is defined as
  \begin{align}
\chi^2=\sum_{i=1}^N\left[\frac{M_i(\alpha,\alpha_0;\beta,\beta_0)-M_i}{M_i}\right]^2.
\end{align}

Inputting the newest data from PDG~\cite{Zyla:2020zbs} as $m_{\eta_c}=2983.9\pm0.5MeV$, $m_{\eta_c(2S)}=3637.5\pm1.1MeV$, $m_{X(3940)}=3942\pm9MeV$, $m_{J/\psi}=3096.9\pm0.006MeV$,
$m_{\psi(2S)}=3686.10\pm0.06MeV$, $m_{\psi(4040)}=4039\pm1MeV$, $m_{\psi(4415)}=4421\pm4MeV$,
$m_{h_{c}}=3525.38\pm0.11MeV$, $m_{\chi_{c2}}=3556.17\pm0.07MeV$,
the fitted parameters for charmonia are
 \begin{align}
&\alpha(\eta_c)=0.35\pm0.04 GeV^{-2},  ~~~ \alpha_0(\eta_c)=-3.17\pm0.43,~~~~\chi^2=0.002,\\
&\alpha(J/\psi)=0.39\pm0.04GeV^{-2},  ~~~ \alpha_0(J/\psi)=-2.86\pm0.50,~~~~\chi^2=0.001,\\
&\beta(\eta_c)=0.29\pm0.03GeV^{-2},  ~~~ \beta_0(\eta_c)=-2.67\pm0.38,~~~~\chi^2=0.004,\\
&\beta(J/\psi)=0.31\pm0.02GeV^{-2},  ~~~ \beta_0(J/\psi)=-3.07\pm0.29,~~~~\chi^2=0.002.
\end{align}
Similarly, we can get the fit results for bottomonia and $B_c$ mesons.
 \begin{align}
 &\alpha(B_c)=0.20\pm0.02,  ~~~ \alpha_0(B_c)=-7.79\pm0.89,~~~~\chi^2=0.001,\\
 &\beta(B_c)=0.15\pm0.02,  ~~~ \beta_0(B_c)=-6.07\pm0.72,~~~~\chi^2=0.001,\\
&\alpha(\eta_b)=0.13\pm0.01,  ~~~ \alpha_0(\eta_b)=-11.43\pm1.27,~~~~\chi^2=0.001,\\
&\alpha(\Upsilon)=0.14\pm0.01,  ~~~ \alpha_0(\Upsilon)=-11.58\pm1.38,~~~~\chi^2=0.001,\\
&\beta(\eta_b)=0.11\pm0.01,  ~~~ \beta_0(\eta_b)=-9.75\pm0.95,~~~~\chi^2=0.001,\\
&\beta(\Upsilon)=0.13\pm0.01,  ~~~ \beta_0(\Upsilon)=-11.66\pm0.84,~~~~\chi^2=0.005.
\end{align}

 \begin{figure}[th]
\begin{center}
\includegraphics[width=0.7\textwidth]{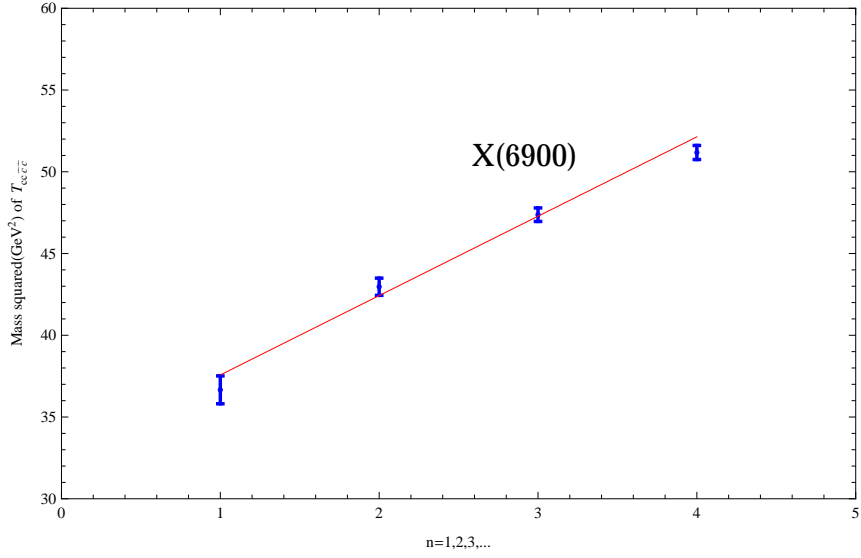}
\caption{ Regge trajectory for fully charm tetraquark $T_{cc\bar{c}\bar{c}}$ with the spin-parity $J^{PC}=0^{++}$.
The $X(6900)$  state observed by LHCb Collaboration may be assigned as a $0^{++}(3S)$ $T_{cc\bar{c}\bar{c}}$ state.}\label{fig:Tcccc}
\end{center}
\end{figure}

\begin{figure}[th]
\begin{center}
\includegraphics[width=0.7\textwidth]{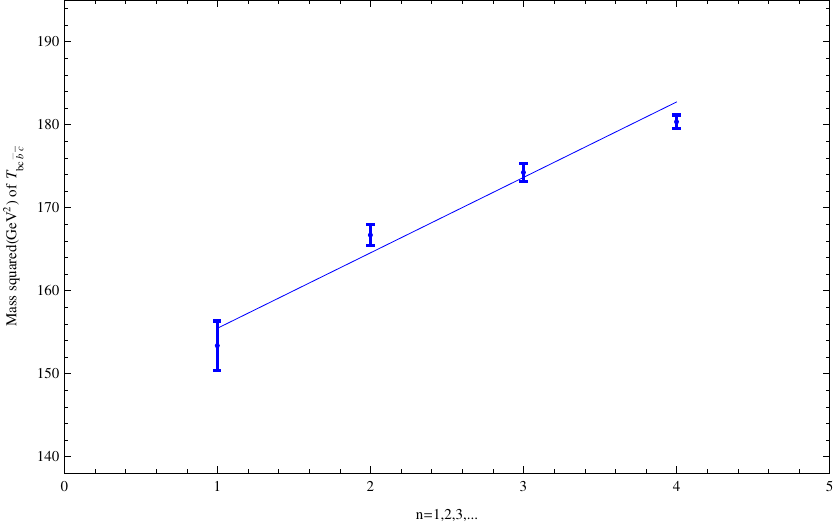}
\caption{ Regge trajectory for fully charm tetraquark $T_{bc\bar{b}\bar{c}}$ with the spin-parity $J^{PC}=0^{++}$.
 }\label{fig:Tbcbc}
\end{center}
\end{figure}

\begin{figure}[th]
\begin{center}
\includegraphics[width=0.7\textwidth]{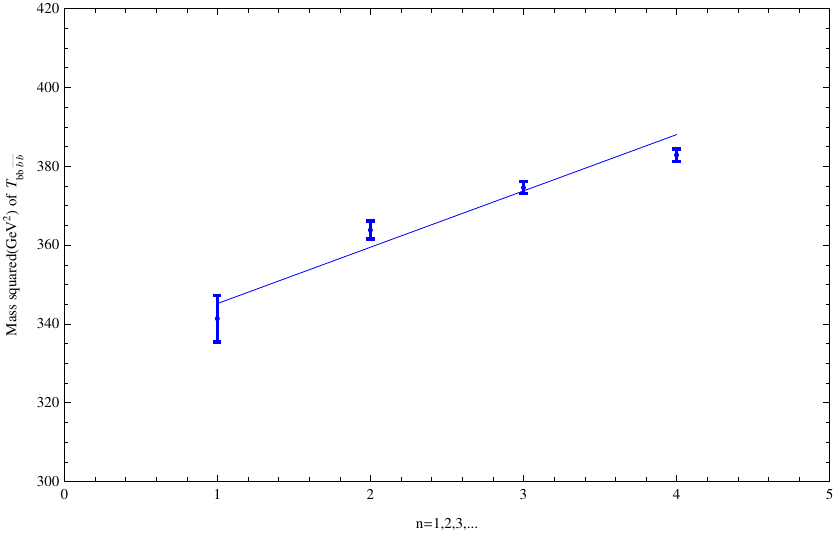}
\caption{ Regge trajectory for fully charm tetraquark $T_{bb\bar{b}\bar{b}}$ with the
spin-parity $J^{PC}=0^{++}$.
 }\label{fig:Tbbbb}
\end{center}
\end{figure}

Actually, we can estimate the  intercepts using the Eq.~(\ref{intercep}). For example,
 \begin{align}
 &\alpha_0(\eta_c)\approx -\alpha(\eta_c)(2m_c)^2\approx-3.11,  ~~~ \beta_0(\eta_c)\approx -\beta(\eta_c)(2m_c)^2\approx-2.59,\\
&\alpha_0(J/\psi)\approx -\alpha(J/\psi)(2m_c)^2+1\approx-2.51,  ~~~ \beta_0(J/\psi)\approx -\beta(J/\psi)(2m_c)^2\approx-2.79,
\end{align}
where the charm quark mass is adopted as $m_c=1.5GeV$~\cite{Zhu:2017lwi,Zhu:2017lqu,Qiao:2012hp,Qiao:2012vt}. Besides,
the slopes $\alpha$ and  $\beta$  are around the strong coupling constant. These estimations  sometimes
are useful when we have little information about the hadron masses.

In the following, we can extract the related parameters for the fully heavy tetraquarks
in Regge trajectories. Using the masses of the fully heavy tetraquarks calculated in the previous section,
the related slopes and intercepts are determined as
 \begin{align}
&\beta(T_{cc\bar{c}\bar{c}}(0^{++}))=0.206\pm0.013,~~~ \beta_0(T_{cc\bar{c}\bar{c}}(0^{++}))=-7.74\pm0.56,\\
&\beta(T'_{cc\bar{c}\bar{c}}(0^{++}))=0.212\pm0.013,~~~ \beta_0(T'_{cc\bar{c}\bar{c}}(0^{++}))=-7.76\pm0.57,\\
&\beta(T_{cc\bar{c}\bar{c}}(2^{++}))=0.211\pm0.013,~~~ \beta_0(T_{cc\bar{c}\bar{c}}(2^{++}))=-8.00\pm0.58,
\end{align}
 \begin{align}
&\beta(T_{bc\bar{b}\bar{c}}(0^{++}))=0.110\pm0.007,~~~ \beta_0(T_{bc\bar{b}\bar{c}}(0^{++}))=-17.10\pm1.15,\\
&\beta(T'_{bc\bar{b}\bar{c}}(0^{++}))=0.112\pm0.007,~~~ \beta_0(T'_{bc\bar{b}\bar{c}}(0^{++}))=-17.14\pm1.15,\\
&\beta(T_{bc\bar{b}\bar{c}}(2^{++}))=0.111\pm0.007,~~~ \beta_0(T_{bc\bar{b}\bar{c}}(2^{++}))=-17.35\pm1.16,
\end{align}
 \begin{align}
&\beta(T_{bb\bar{b}\bar{b}}(0^{++}))=0.070\pm0.004,~~~ \beta_0(T_{bb\bar{b}\bar{b}}(0^{++}))=-24.16\pm1.61,\\
&\beta(T'_{bb\bar{b}\bar{b}}(0^{++}))=0.071\pm0.004,~~~ \beta_0(T'_{bb\bar{b}\bar{b}}(0^{++}))=-24.17\pm1.61,\\
&\beta(T_{bb\bar{b}\bar{b}}(2^{++}))=0.070\pm0.004,~~~ \beta_0(T_{bb\bar{b}\bar{b}}(2^{++}))=-24.33\pm1.62.
\end{align}
 As examples, we plot the
$(n_r, M^2)$ plane Regge trajectories of $T_{cc\bar{c}\bar{c}}(0^{++}))$, $T_{bc\bar{b}\bar{c}}(0^{++}))$, and $T_{bb\bar{b}\bar{b}}(0^{++}))$  in Fig.~\ref{fig:Tcccc}, Fig.~\ref{fig:Tbcbc}, and Fig.~\ref{fig:Tbbbb}, respectively.  The LHCb $X(6900)$ state may be assigned as a $0^{++}(3S)$ $T_{cc\bar{c}\bar{c}}$ state. However, it is also possible to assign it as $0^{++}(3S)$ $T'_{cc\bar{c}\bar{c}}$ state or $2^{++}(3S)$ $T_{cc\bar{c}\bar{c}}$ state or an orbitally $2P$ state. To further determine its nature, one may need to
investigate its decay width or production cross section.

\section{Production at hadron colliders}
\label{SEC:production}
The cross section of the fully heavy tetraquark at proton-proton collider shall be factorized as

\begin{align}
\sigma\left(p+p\to T_{4Q}+X\right)=& \sum_{i,j=q,g} \int_{0}^{1} d x_{1} d x_{2} f_{i / p}\left(x_{1}, \mu\right)
f_{j / p}\left(x_{2}, \mu\right) \int_{0}^{1} d z \hat{\sigma}_{ij}^{(0)} \nonumber\\&\times H_{ij}\left(z ;\mu\right)\delta\left(z-\frac{m^2_{T_{4Q}}}{x_{1} x_{2}s}\right),
\end{align}
where $T_{4Q}$ denotes one of the fully heavy tetraquarks $T_{cc\bar{c}\bar{c}}$, $T_{bb\bar{b}\bar{b}}$ and $T_{bc\bar{b}\bar{c}}$; $\hat{\sigma}_{ij}^{(0)}$ is
the LO cross section for the partonic subprocess $i+j\to T_{4Q}+X$; $H_{ij}$ is the hard kernel; $x_i$ is the parton longitudinal
momentum fraction and $s$ is the centre-of-mass
energy of incoming protons.
To produce the fully heavy tetraquarks, two pair of heavy quarks should be created at first.
Thus the two gluon fusion is the dominant production mechanism for the fully heavy tetraquarks. Up to NLO,
the following processes should be considered
$$ p+p\to g+g \to T_{4Q},~~~  p+p\to g+g \to T_{4Q}+g. $$

The hard kernel $H_{ij}$ can be expanded in powers of strong coupling constant
\begin{align}
H_{ij}\left(z ;\mu\right)=&\sum_n \left(\frac{\alpha_s}{2\pi}\right)^n H^{(n)}_{ij}\left(z ;\mu\right),\\
H^{(0)}_{ij}\left(z ;\mu\right)=&\delta_{ig}\delta_{jg}\delta(1-z).
\end{align}
\begin{figure}[th]
\begin{center}
\includegraphics[width=0.7\textwidth]{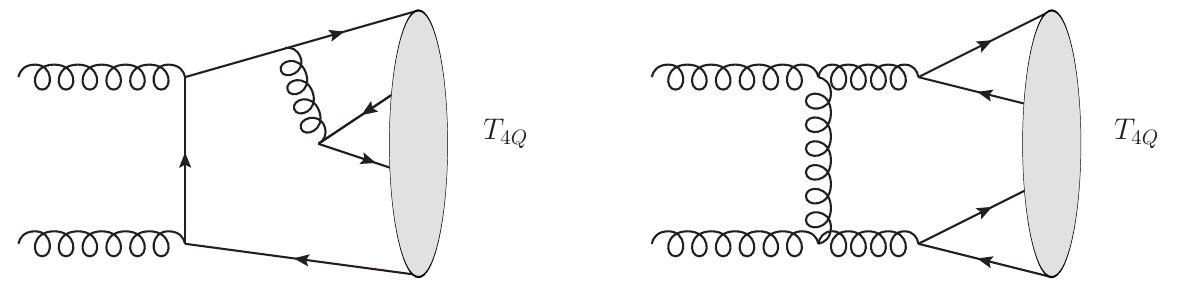}
\caption{ Typical Feynman diagrams for the production of fully heavy tetraquark $T_{4Q}$.
 }\label{fig:T4Q}
\end{center}
\end{figure}

The typical Feynman diagrams for $g+g \to T_{4Q}$ are plotted in Fig.~\ref{fig:T4Q}. The LO partonic cross section $\hat{\sigma}_{gg}^{(0)}$
are related to the LO Feynman amplitude squared
\begin{align}
\hat{\sigma}_{gg}^{(0)}=&\frac{\pi}{(D-2)^2(N_c^2-1)^2}|{\cal M}(g+g \to T_{4Q})|^2.
\end{align}
In this paper, we only consider the S-wave tetraquark production. It is convenient to write the partonic  amplitude into Lorentz invariant terms
\begin{align}
{\cal M}[g(\epsilon_1(p_1))+g(\epsilon_2(p_2)) \to T_{4Q}(0^{++}, p_H)]=&
\left[a g_{\mu\nu}+b \frac{{p_H}_\mu {p_H}_\nu}{p_H^2}\right]\epsilon_1^\mu\epsilon_2^\nu,\nonumber\\
{\cal M}[g(\epsilon_1(p_1))+g(\epsilon_2(p_2)) \to T_{4Q}(2^{++}, \epsilon^*(p_H))]=&
\left[c \epsilon^*_{\mu\nu}+d \frac{{p_H}_\mu {p_H}_\nu {p_1}_\alpha {p_1}_\beta \epsilon^*_{\alpha\beta}}{p_H^4}
+f \frac{g_{\mu\nu} {p_1}_\alpha {p_1}_\beta \epsilon^*_{\alpha\beta}}{p_H^2}
\right.\nonumber\\&\left.+g \frac{{p_H}_\nu {p_1}_\alpha \epsilon^*_{\mu\alpha}}{p_H^2}
+h \frac{{p_H}_\mu {p_1}_\alpha \epsilon^*_{\nu\alpha}}{p_H^2}\right] \epsilon_1^\mu\epsilon_2^\nu.\nonumber
\end{align}
It is a hard task to calculate the four body production matrix elements for fully heavy tetraquarks. For an {\it ab initio} method,
we will employ the NRQCD to simplify the LDMEs for fully heavy tetraquarks as the series of  two-body LDMEs
\begin{align}
\langle 0|\mathcal{O}^{T_{4Q}}|0\rangle=&\sum_i c_{\mathbf{1}i}\frac{\langle 0|\mathcal{O}^{Q\bar{Q}[^{2S+1}L_J]^{[\mathbf{1}i]}}|0\rangle\langle 0
|\mathcal{O}^{Q\bar{Q}[^{2S+1}L_J]^{[\mathbf{1}i]}}|0\rangle}{m^2_{T_{4Q}}}
\nonumber\\&+\sum_j c_{\mathbf{8}j}\frac{\langle 0|\mathcal{O}^{Q\bar{Q}[^{2S+1}L_J]^{[\mathbf{8}i]}}|0\rangle\langle 0
|\mathcal{O}^{Q\bar{Q}[^{2S+1}L_J]^{[\mathbf{8}i]}}|0\rangle}{m^2_{T_{4Q}}},
\end{align}
where the number ${\bf 1}$ and ${\bf 8}$ denote the color singlet and octet. By Fierz transformation, the above decomposition
can be performed in a diquark and anti-diquark configurations ${\bf 3} \otimes  {\bf\bar 3}$ and ${\bf 6} \otimes  {\bf\bar 6}$.
One can see the decomposition of a diquark and anti-diquark configurations in Ref.~\cite{Feng:2020riv}.

Since the color-octet LDMEs of heavy quarkonium are small,  we just consider the color-singlet contribution here. Since
$0^{++}$ can be produced by vector-vector and pseudoscalar-pseudoscalar configurations, while $2^{++}$ can be produced by
vector-vector configuration. Thus we denote the coefficient of vector-vector coupling to the tetraquark LDMEs is denoted as
$c_{\mathbf{1}1}$ and the coefficient of pseudoscalar-pseudoscalar coupling to the tetraquark LDMEs is denoted as
$c_{\mathbf{1}0}$. We leave a complete investigation of all other possible LDMEs contributions in future works. Then the LO partonic cross sections are

\begin{align}
\hat{\sigma}_{gg}^{(0)}(T_{bc\bar{b}\bar{c}}(0^{++}))=&
\frac{4 \pi ^5 (r+1)^8 \alpha _s^4 \left(C_A-2 C_F\right){}^2}{27 s_J^2r^4  m_{T_{bc\bar{b}\bar{c}}}^8}c_{\mathbf{1}1}\left[\langle 0|\mathcal{O}^{b\bar{c}}
 (^{3}S_{1}^{[1]}))|0\rangle\right]^2,
\end{align}
\begin{align}
\hat{\sigma}_{gg}^{(0)}(T'_{bc\bar{b}\bar{c}}(0^{++}))=& \frac{4 \pi ^5 (r+1)^8 \alpha _s^4
\left(C_A-10 C_F\right){}^2}{81 r^4  m_{T_{bc\bar{b}\bar{c}}}^8}c_{\mathbf{1}0}\left[\langle 0|\mathcal{O}^{b\bar{c}}
 (^{1}S_{1}^{[0]}))|0\rangle\right]^2,
\end{align}
\begin{align}
\hat{\sigma}_{gg}^{(0)}(T_{bc\bar{b}\bar{c}}(2^{++}))=&
\frac{64 \pi ^5  (r+1)^8 \alpha _s^4 \left(C_A-4 C_F\right){}^2}{81 s_J^2 r^4  m_{T_{bc\bar{b}\bar{c}}}^8}c_{\mathbf{1}1}\left[\langle 0|\mathcal{O}^{b\bar{c}}
 (^{3}S_{1}^{[1]}))|0\rangle\right]^2,
\end{align}
where $r=m_c/m_b$ and $s_J=3$. One can easily get the LO partonic cross sections for fully charm or bottom tetraquarks.
$\hat{\sigma}_{gg}^{(0)}(T_{cc\bar{c}\bar{c}}(0^{++}, 2^{++}))$
can be obtained by the replacement $m_{T_{bc\bar{b}\bar{c}}}\to m_{T_{cc\bar{c}\bar{c}}}$, $\langle 0|\mathcal{O}^{b\bar{c}}
 (^{3}S_{1}^{[1]}))|0\rangle \to \langle 0|\mathcal{O}^{c\bar{c}}
 (^{3}S_{1}^{[1]}))|0\rangle$ ( or $ \langle 0|\mathcal{O}^{c\bar{c}}
 (^{1}S_{0}^{[1]}))|0\rangle$), and $r\to 1$.

Using the LDMEs of S-wave charmonium, bottomonium, and $B_c$ meson in Refs.~\cite{Zhu:2017lqu,Zhu:2015qoa,Zhu:2015jha,Zhu:2018bwp},
the fully heavy tetraquark hadroproduction
cross section can be obtained as

\begin{align}
\sigma(X(6900),T_{cc\bar{c}\bar{c}}(0^{++}))&=
c_{\mathbf{1}1}\left\{ \begin{array} {ll}(9.4,18.2) {\rm nb}\ , & \sqrt{s}= 2.75{\rm TeV}\ ,\\
(21.4,41.5) {\rm nb}\ , & \sqrt{s}= 7{\rm TeV}\ , \\ (37.2,72.3) {\rm nb}\ , & \sqrt{s}= 14{\rm TeV}\ , \end{array} \right.
 \end{align}
 \begin{align}
\sigma(X(6900),T'_{cc\bar{c}\bar{c}}(0^{++}))&=
c_{\mathbf{1}0}\left\{ \begin{array} {ll}(3007,5840) {\rm nb}\ , & \sqrt{s}= 2.75{\rm TeV}\ ,\\
(6845,13293) {\rm nb}\ , & \sqrt{s}= 7{\rm TeV}\ , \\ (12097,23494) {\rm nb}\ , & \sqrt{s}= 14{\rm TeV}\ , \end{array} \right.
 \end{align}
 \begin{align}
\sigma(X(6900),T_{cc\bar{c}\bar{c}}(2^{++}))&=
c_{\mathbf{1}1}\left\{ \begin{array} {ll}(2453,4764) {\rm nb}\ , & \sqrt{s}= 2.75{\rm TeV}\ ,\\
(5584,10845) {\rm nb}\ , & \sqrt{s}= 7{\rm TeV}\ , \\ (9729,18895) {\rm nb}\ , & \sqrt{s}= 14{\rm TeV}\ , \end{array} \right.
 \end{align}
 \begin{align}
\sigma(T_{bc\bar{b}\bar{c}}(0^{++}))&=
c_{\mathbf{1}1}\left\{ \begin{array} {ll}(0.09,0.16) {\rm nb}\ , & \sqrt{s}= 2.75{\rm TeV}\ ,\\
(0.25,0.44) {\rm nb}\ , & \sqrt{s}= 7{\rm TeV}\ , \\ (0.49,0.86) {\rm nb}\ , & \sqrt{s}= 14{\rm TeV}\ , \end{array} \right.
 \end{align}
  \begin{align}
\sigma(T'_{bc\bar{b}\bar{c}}(0^{++}))&=
c_{\mathbf{1}0}\left\{ \begin{array} {ll}(28.8,50.9) {\rm nb}\ , & \sqrt{s}= 2.75{\rm TeV}\ ,\\
(79.3,140.0) {\rm nb}\ , & \sqrt{s}= 7{\rm TeV}\ , \\ (155.4,274.5) {\rm nb}\ , & \sqrt{s}= 14{\rm TeV}\ , \end{array} \right.
 \end{align}
 \begin{align}
\sigma(T_{bc\bar{b}\bar{c}}(2^{++}))&=
c_{\mathbf{1}1}\left\{ \begin{array} {ll}(23.5,41.6) {\rm nb}\ , & \sqrt{s}= 2.75{\rm TeV}\ ,\\
(64.7,114.2) {\rm nb}\ , & \sqrt{s}= 7{\rm TeV}\ , \\ (126.8,223.9) {\rm nb}\ , & \sqrt{s}= 14{\rm TeV}\ , \end{array} \right.
 \end{align}
  \begin{align}
\sigma(T_{bb\bar{b}\bar{b}}(0^{++}))&=
c_{\mathbf{1}1}\left\{ \begin{array} {ll}(0.03,0.05) {\rm nb}\ , & \sqrt{s}= 2.75{\rm TeV}\ ,\\
(0.09,0.14) {\rm nb}\ , & \sqrt{s}= 7{\rm TeV}\ , \\ (0.18,0.30) {\rm nb}\ , & \sqrt{s}= 14{\rm TeV}\ , \end{array} \right.
 \end{align}
  \begin{align}
\sigma(T'_{bb\bar{b}\bar{b}}(0^{++}))&=
c_{\mathbf{1}0}\left\{ \begin{array} {ll}(8.7,14.7) {\rm nb}\ , & \sqrt{s}= 2.75{\rm TeV}\ ,\\
(27.3,46.0) {\rm nb}\ , & \sqrt{s}= 7{\rm TeV}\ , \\ (57.7,97.3) {\rm nb}\ , & \sqrt{s}= 14{\rm TeV}\ , \end{array} \right.
 \end{align}
 \begin{align}
\sigma(T_{bb\bar{b}\bar{b}}(2^{++}))&=
c_{\mathbf{1}1}\left\{ \begin{array} {ll}(7.1,12.0) {\rm nb}\ , & \sqrt{s}= 2.75{\rm TeV}\ ,\\
(22.2,37.6) {\rm nb}\ , & \sqrt{s}= 7{\rm TeV}\ , \\ (79.3,147.0) {\rm nb}\ , & \sqrt{s}= 14{\rm TeV}\ , \end{array} \right.
 \end{align}
 where the scale is adopted at $(2m_{T_{4Q}}, m_{T_{4Q}})$. The coefficients $c_{\mathbf{1}0}$ and $c_{\mathbf{1}1}$
 are not determined, however, one can estimate their magnitude. In Ref.~\cite{Chatrchyan:2013cld}, the CMS collaboration
 have measured the product of the cross section of $X(3872)$ and its branching fraction into $J/\psi \pi^+\pi^-$ at $\sqrt{s}= 7{\rm TeV}$
  as $\sigma(X(3872))Br(X(3872)\to J/\psi \pi^+\pi^-)=1.06\pm0.11\pm0.15{\rm nb}$.
  The bounds on the cross section of $X(3872)$ was extracted as $2.6{\rm nb}<\sigma(X(3872))<31{\rm nb}$ in Ref.~\cite{Braaten:2018eov}.
  In Ref.~\cite{Carvalho:2015nqf}, the  $X(3872)$ was treated as a $c\bar{c}q\bar{q}$ tetraquark and its cross section was
  predicted as $10{\rm nb}<\sigma(X(3872))<40{\rm nb}$ at $\sqrt{s}= 7{\rm TeV}$.  The cross section of fully charm tetraquark
  $T_{cc\bar{c}\bar{c}}$  was
  also predicted in Ref.~\cite{Carvalho:2015nqf} as $3.6\pm2.5 {\rm nb}$ at $\sqrt{s}= 7{\rm TeV}$.  Then one can
 obtain the above limits for the coefficients as $c_{\mathbf{1}1}\leq 10^{-3}$ and $c_{\mathbf{1}0}\leq 10^{-3}-10^{-4}$.
 Considering the origin of the wave function approaches as $\chi_p(r\to 0)\propto -4\times 10^{-4}$ for $T_{cc\bar{c}\bar{c}}$ in Sec. II,
 the coefficients are determined as $c_{\mathbf{1}1}\sim 10^{-5}-10^{-6}$ and $c_{\mathbf{1}0}\sim 10^{-5}-10^{-6}$. Then the cross section of the $X(6900)$
 is estimated around $(10-100)fb$.

 Recently the LHCb collaboration have measured the production cross section of double $J/\psi$ as $15.2\pm1.0\pm0.9nb$ at 13TeV~\cite{Aaij:2016bqq}. The ATLAS collaboration also measured the double parton scattering contributions
 in  double $J/\psi$ channels~\cite{Aaboud:2016fzt}. If we only consider the single parton scatting processes, the $X(6900)$ signal/background is around $1/100-1/1000$ for the LHCb experiment.

From the above calculation, the cross section of a $2^{++}$ tetraquark is close to that of  a $0^{++}$ tetraquark from
pseudoscalar-pseudoscalar configuration, both of which are greatly larger than the cross section of
a $0^{++}$ tetraquark from vector-vector configuration, which is important to determine the nature of $X(6900)$
if the $X(6900)$  is a S-wave tetraquark. Furthermore, we have
\begin{equation}
\frac{\sigma(T_{cc\bar{c}\bar{c}}(2^{++}))}{\sigma(T_{cc\bar{c}\bar{c}}(0^{++}))}\sim 260,
\end{equation}
  \begin{equation}
\frac{\sigma(T'_{cc\bar{c}\bar{c}}(0^{++}))}{\sigma(T_{cc\bar{c}\bar{c}}(0^{++}))}\sim 320,
\end{equation}
 \begin{equation}
\frac{\sigma(T_{cc\bar{c}\bar{c}}(2^{++}(2S)))}{\sigma(T_{cc\bar{c}\bar{c}}(2^{++}(1S)))}\sim 0.6,
\end{equation}
\begin{equation}
\frac{\sigma(T_{cc\bar{c}\bar{c}}(2^{++}(3S)))}{\sigma(T_{cc\bar{c}\bar{c}}(2^{++}(1S)))}\sim 0.4.
\end{equation}

The cross section of tetraquark is scaled as $1/\sqrt{s}$, which is another phenomenon to test the theoretical method.
The cross section of $T_{bc\bar{b}\bar{c}}$ is one percent of that of $T_{cc\bar{c}\bar{c}}$, while the cross section
of $T_{bb\bar{b}\bar{b}}$ is suppressed by a factor $1/300$ compared to the cross section of $T_{cc\bar{c}\bar{c}}$.

To hunt for the $T_{bc\bar{b}\bar{c}}$  states, one could study the process $p+p\to g+g\to T_{bc\bar{b}\bar{c}}$.
For a $T_{bc\bar{b}\bar{c}}$  state around 12.4GeV, one could use the decay channel
 $T_{bc\bar{b}\bar{c}}\to \Upsilon+\ell^-+\ell^+$ or  $T_{bc\bar{b}\bar{c}}\to J/\psi+\ell^-+\ell^+$.
 For a $T_{bc\bar{b}\bar{c}}$  state around 12.9GeV or 13.2GeV, one could use the decay channel
 $T_{bc\bar{b}\bar{c}}\to \Upsilon+J/\psi$.

 To hunt for the $T_{bc\bar{b}\bar{c}}$  states, one could also study the process $p+p\to g+g\to T_{bb\bar{b}\bar{b}}$.
For a $T_{bb\bar{b}\bar{b}}$  state around 18.5GeV, one could use the decay channel
 $T_{bb\bar{b}\bar{b}}\to \Upsilon+\ell^-+\ell^+$.
 For a $T_{bb\bar{b}\bar{b}}$  state around 19.1GeV 0r 19.4GeV, one could use the decay channel
 $T_{bb\bar{b}\bar{b}}\to \Upsilon+\Upsilon$.

\section{Differential cross section at low transverse momentum}
\label{SEC:pt}

Concerning about the differential cross section, the LO Feyman diagrams only give a delta function.
We need to consider
the process $g+g\to T_{4Q}+g$.  But
we can study its behaviour at low transverse momentum limit. In the low transverse momentum limit $p_\perp\ll m_{T_{4Q}}$,
 the differential cross section becomes
\begin{align}
\frac{d \sigma}{d y d^{2} p_{\perp}} &=\hat{\sigma}_{gg}^{(0)} \frac{\alpha_{s} C_{A}}{2 \pi^{2}} \int dx_1 dx_2 f(x_1,\mu)
 f\left(x_2,\mu\right)  \frac{1}{p_{\perp}^{2}}\left[\frac{2\left(1-\xi_{1}+\xi_{1}^{2}\right)^{2}}{\left(1-\xi_{1}\right)_{+}}
 \delta\left(1-\xi_{2}\right)\right.\nonumber\\
&\left.+\frac{2\left(1-\xi_{2}+\xi_{2}^{2}\right)^{2}}{\left(1-\xi_{2}\right)_{+}} \delta\left(1-\xi_{1}\right)
+2 \ln \frac{{m^2_{T_{4Q}}}}{p_{\perp}^{2}} \delta\left(1-\xi_{2}\right) \delta\left(1-\xi_{1}\right)\right],
\end{align}
where $y$ is the rapidity; $p_\perp$ is the transverse momentum of tetraquark; $\xi_1=m_{T_{4Q}} e^y/(x_1\sqrt{s})$,
and $\xi_2=m_{T_{4Q}} e^y/(x_2\sqrt{s})$. This formalism will break down when $p_\perp \to 0$. Thus we use the
Collins-Soper-Sterman resummation formula~\cite{Collins:1984kg} and the differential cross section
can be rewritten as~\cite{Sun:2012vc,Zhu:2013yxa}
\begin{align}
\left.\frac{d \sigma}{dy d^{2} p_{\perp} }\right|_{p_{\perp} \ll m_{T_{4Q}}}=\frac{1}{(2 \pi)^{2}}
\int d^{2} b e^{i \vec{p}_{\perp} \cdot \vec{b}} e^{-{\cal S}_{sud}(b,m_{T_{4Q}},C_1,C_2)}W\left(b, m_{T_{4Q}},\xi_1,\xi_2\right),
\end{align}is
where ${\cal S}_{sud}(m_{T_{4Q}},C_1,C_2)$ is the Sudakov factor
\begin{align}
{\cal S}_{sud}(b,m_{T_{4Q}},C_1,C_2)=\int_{C^2_1/b^2}^{C^2_2m^2_{T_{4Q}}}\frac{d\mu^2}{\mu^2}\left[A\log\frac{C^2_2m^2_{T_{4Q}}}{\mu^2}
+B\right],
\end{align}
where both $A$ and $B$ can be expanded perturbatively as $A( B)=\sum_i \left(\frac{\alpha_s}{2\pi} \right)^i A^{(i)}(B^{(i)})$.
For the lowest nontrivial order, $A^{(1)}=2C_A$ and $B^{(1)}=-2b_0=-(11C_A/3-2n_f/3)$. It is popular to choose $C_1=2e^{-\gamma_E}$ and $C_2=1$.

$W\left(b, m_{T_{4Q}},\xi_1,\xi_2\right)$ can be written as
\begin{align}
W\left(b, m_{T_{4Q}},\xi_1,\xi_2\right)=& \hat{\sigma}_{gg}^{(0)}  \int dx_1 dx_2 f_a(x_1,\mu)
 f_b\left(x_2,\mu\right)  C_{g a}\left(\frac{\xi'_{1}}{x_1}, b, C_{1}, C_{2}, \mu^{2}\right) \nonumber\\
& \times C_{g b}\left(\frac{\xi'_{2}}{x_2}, b, C_{1}, C_{2}, \mu^{2}\right),
\end{align}
where $\xi'_1=m_{T_{4Q}} e^y/\sqrt{s}$,
and $\xi'_2=m_{T_{4Q}} e^y/\sqrt{s}$.  $C_{ij}$ rely  on the fixed perturbative calculation and
can be expanded as $C_{ij}=\sum_n \left(\frac{\alpha_s}{2\pi} \right)^n C_{ij}^{(n)}$, and at leading order,
 $C_{gg}^{(0)}(x)=\delta(1-x)$ and $C_{gq}^{(0)}(x)=0$.
We will leave the higher-order QCD corrections in future studies.

\begin{figure}[th]
\begin{center}
\includegraphics[width=0.7\textwidth]{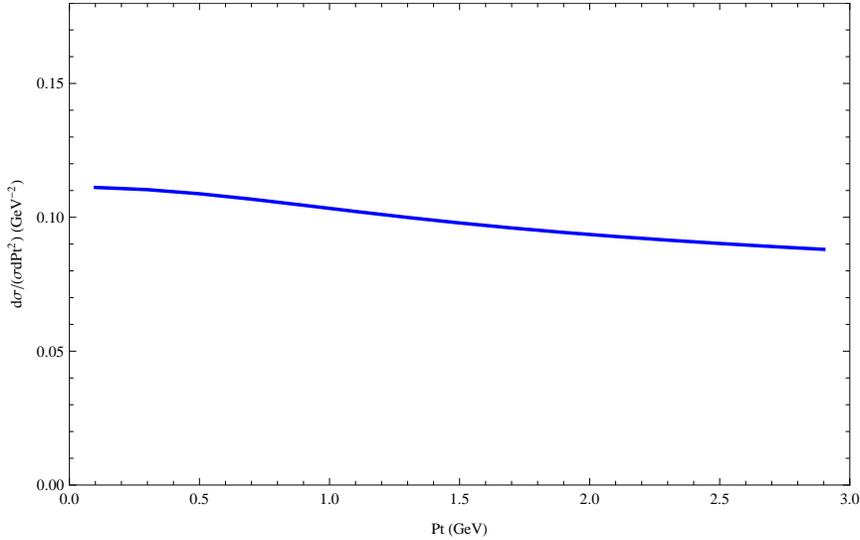}
\caption{ The $X(6900)$ production with $\sqrt{s}=14$TeV and $2.5<y<5$ at the LHC. Pt=$p_\perp$ is the transverse momentum
of the  $X(6900)$.
 }\label{fig:pt}
\end{center}
\end{figure}

For the $X(6900)$ production at proton-proton collision,
we give a plot for its differential cross section $d\sigma/(\sigma  d P_t^2)$  at low transverse momentum in Fig.~\ref{fig:pt}.
After resummation, the cross section will not break down near zero transverse momentum.

\section{Summary}
\label{SEC:conclusion}
We have presented an analyse of the spectra of fully heavy tetraquarks within Bethe-Salpeter equation and Regge trajectories
and a calculation for the production of fully heavy tetraquarks at hadron colliders.
The $X(6900)$ discovered by the LHCb collaboration could be explained as a radially excited S-wave fully heavy
tetraquark  or a orbitally excited P-wave tetraquark. The $X(6900)$ discovery indicates that
the existence of fully heavy tetraquark partners, $T_{cc\bar{c}\bar{c}}$, $T_{bc\bar{b}\bar{c}}$
and $T_{bb\bar{b}\bar{b}}$. The production of both $T_{bc\bar{b}\bar{c}}$
and $T_{bb\bar{b}\bar{b}}$ have a suppression factor, however, these heavier states shall be
tested within a larger data sample of proton-proton collision.

\acknowledgments
The author thanks the useful discussions with Chao-Hsi Chang,
Fernando Silveira Navarra, Cong-Feng Qiao, Peng Sun, Xiangpeng Wang and Kai Yi.
This work is supported by NSFC under grant No.~11705092 and
 12075124, and by Natural Science Foundation of Jiangsu under Grant No.~BK20171471
  and Jiangsu Qinglan project.

\end{document}